\documentclass[12pt]{article}
\begin{document}
\begin{center}
 \textbf{Reply to Comment by J.A.Garcia,  arXiv:0705.0143 (to appear in PRD)} \vskip 2cm

\textbf{Subir Ghosh}\\
Physics and Applied Mathematics Unit,\\
Indian Statistical Institute,\\
203 B. T. Road, Calcutta 700108, India.
\end{center}
\vskip 3cm
{\bf Abstract:}\\
We observe that there is no clash between the works \cite{gar} and \cite{g1}.
\vskip 1cm
In the Comment \cite{gar} the author has shown that one can construct a Lagrangian model of a point particle with a Magueijo-Smolin (MS) form of dispersion relation in a  canonical phase space provided one modifies the Lorentz generator to
\begin{equation}
J^{\mu\nu}_{DSR}=(x^\mu -\frac{(xp)}{l}\eta ^\mu )p^{\nu}-(x^\nu -\frac{(xp)}{l}\eta ^\nu )p^{\mu}.
\label{1}
\end{equation}
On the other hand in \cite{g1} I have shown that one can keep the Lorentz generator 
\begin{equation}
J^{\mu\nu}=x^\mu p^{\nu}-x^\nu p^{\mu}
\label{2}
\end{equation}
 unchanged provided a non-canonical symplectic structure is used.

In my opinion the above two formalisms are complimentary and there is no reason to treat the former \cite{gar} as an improvement, but for a bias of the author of \cite{gar} against the introduction of a non-canonical phase space. 

Furthermore, it is crucial to keep in mind that from the point of view of DSR, the Lagrangian in (4) \cite{gar} with the chosen form of $J^{\mu\nu}_{DSR}$ is fundamental and the coordinate $x^\mu$, (with its non-canonical behavior under Lorentz transformation), is the {\it{physical}} coordinate and $p^{\mu}$ is the {\it{physical}} momentum. According to DSR, results obtained in $x^\mu ,p^\mu $ variables (and {\it{not}} in $X^\mu ,P^\mu $) should be compared with experiments. Thus (4) in \cite{gar} should be considered as the starting point and (1) in \cite{gar} is  obtained in a particular parameterization. This does not mean that the DSR model is trivially related to normal particle model. This is because in order to get the correct behavior of a DSR particle one has to convert the normal particle equations (in $X^\mu ,P^\mu $) to equations involving physical DSR coordinates $x^\mu ,p^\mu $ using
$$X^\mu =(F^{-1})^{\mu}_{\nu}x^\nu $$
as given in \cite{gar}. This will lead to new $\kappa$-DSR physics since coordinates and momenta get mixed up under Lorentz transformation. In this way one can exploit the canonical framework to generate DSR behavior. This sort of approach is discussed extensivly in \cite{g2} in a related context where it is also shown that dynamical inputs are required in order to extrapolate kinematical equations in canonical framework to equations in DSR framework.

\end{document}